\documentclass[prd,a4paper,showpacs,showkeys,preprint,byrevtex]{revtex4}
\usepackage{graphicx}
\usepackage{dcolumn}
\usepackage{amsmath}
\usepackage{bm}

\begin{document}

%\preprint{Preprint Universit\'{e} de Mons-Hainaut}
%\draft
\title{Auxiliary fields and hadron dynamics}

\author{Claude \surname{Semay}}
\thanks{FNRS Research Associate}
\email[E-mail: ]{claude.semay@umh.ac.be}
\affiliation{Service de Physique G\'{e}n\'{e}rale et de Physique des
Particules \'{E}l\'{e}mentaires, Groupe de Physique Nucl\'{e}aire
Th\'{e}orique,
Universit\'{e} de Mons-Hainaut, Place du Parc 20,
B-7000 Mons, Belgium}

\author{Bernard \surname{Silvestre-Brac}}
%\thanks{}
\email[E-mail: ]{silvestre@lpsc.in2p3.fr}
\affiliation{Laboratoire de Physique Subatomique et de Cosmologie,
Avenue des Martyrs 53, F-38026 Grenoble-Cedex, France}

\author{Ilya M. \surname{Narodetskii}}
%\thanks{}
\email[E-mail: ]{naro@heron.itep.ru}
\affiliation{Institute of
Theoretical and Experimental Physics, B. Cheremushkinskaya 25,
RUS-117218 Moscow, Russia}

\date{\today}

\begin{abstract}
The relations existing between the auxiliary field (einbein field)
formalism and the spinless Salpeter equation are studied in the case of
two particles with the same mass, interacting via a confining potential.
The problem of non-orthogonality for radial excited states in the
auxiliary field formalism is discussed and found to be non-crucial. It
is shown that the classical equations of motion of the rotating string
model, derived from the QCD lagrangian, reduce exactly to the classical
equations of motion of the phenomenological semirelativistic flux tube
model, provided all auxiliary fields are eliminated correctly from the
rotating string hamiltonian.
\end{abstract}

% 12.39.Pn    Potential model
% 12.39.Ki    Relativistic quark model
% 14.40.-n    Mesons

\maketitle

\section{Introduction}
\label{sec:intro}

A system of relativistic particles can be quantized as a
constrained system by using the auxiliary field formalism (also
known as einbein field formalism) to get rid of the square root
term in the Lagrangian \cite{polyakov,brin77}. Applied to
the QCD Lagrangian, this technique yields Hamiltonians for mesons
or baryons $H(\mu_i)$ depending on auxiliary fields $\mu_i$
\cite{simonov,kala97,naro02}. The way of removing these fields for
physical applications is not well defined and can lead to
ambiguities. In practice, these fields, representing for instance
the energy density for the particles, are finally treated as
$c$-number and determined from the minimum energy conditions
\begin{equation}
\label{minicond}
\frac{\partial E(\mu_i)}{\partial \mu_i} = 0,
\end{equation}
where $E(\mu_i)$ is an eigenvalue of $H(\mu_i)$.

When the interaction $V$ between particles does not depend on $\mu_i$,
the values of $\mu_i$ which minimize the hamiltonian operator $H(\mu_i)$
(and not $E(\mu_i)$) are $\mu_i = \sqrt{\vec p\,_i^2 + m_i^2}$. The
eigenvalue equation reduces then to a spinless Salpeter equation with
the potential $V$. So, in this case, the auxiliary field formalism can
be considered as an approximation of the spinless Salpeter equation
\cite{luch96}.

Although handling a relativistic expression for the kinetic energy term
is not a serious problem with nowadays numerical algorithms, a
nonrelativistic expression is always simpler, specially for many body
calculations. In a sense the auxiliary field formalism simulates a
relativistic expression with the simplicity of a nonrelativistic one.
The price to pay is a further minimization on eigenenergies. The values
of $\mu_i$ are generally different for various radial excited states and
this leads to non-orthogonality among them. This is another price to
pay. An aim of this paper is to study whether these drawbacks are very
inconvenient or not.

In Sec.~\ref{sec:sse}, we consider the case of two identical
particles interacting via a potential independent of auxiliary fields.
We solve the eigenvalue equations for hamiltonians written in the
auxiliary field formalism and for corresponding hamiltonians in spinless
Salpeter equations. Then, we compare the eigenvalues and
eigenvectors for these two approaches and discuss the problem of
non-orthogonality. The study is first made for a toy model where most of
the
results are analytical and then in a more realistic one which relies on
a precise numerical treatment.

It is now believed that the color field between a quark and an antiquark
in a meson can be approximated by a rigid string, carrying both energy
and angular momentum. Some years ago, a phenomenological model taking
into account this particular string dynamics has been developed: the
relativistic flux tube model \cite{laco89,olss95}. The classical
equations of motion take the form of a coupled system of nonlinear
equations in which a particular quantity is introduced, the transverse
velocity of the ends of the string, where the particles are located.
More recently, a lagrangian for a rotating string was derived from the
QCD lagrangian \cite{dubi94}. Adopting a simple straight line ansatz for
the string, another coupled system of nonlinear equations was obtained,
containing auxiliary functions for the particle energy density and the
string energy density \cite{morg99}. In this case, the potential does
depend on the auxiliary fields. In Sec.~\ref{sec:string}, we
show that these two approaches are exactly the same, provided all
auxiliary functions are correctly eliminated from the rotating string
hamiltonian.

\section{Auxiliary fields and Spinless Salpeter equation}
\label{sec:sse}

In the formalism of auxiliary fields, the hamiltonian $H^A$ for two
particles with the same mass $m$, interacting via a potential $V$
which does not depend on $\mu$, is written \cite{kala97}
($\hbar = c =1$)
\begin{equation}
\label{hn}
H^A=\frac{\vec p\,^2 + m^2}{\mu} + \mu +V.
\end{equation}
The philosophy of this formalism is to keep the simplicity of the
nonrelativistic form of the kinetic energy
operator, to get the eigenenergies in terms of auxiliary fields, and to
get rid of them through a minimization procedure.

As mentioned, the eigenvalues and eigenvectors of this hamiltonian
depends on $\mu$
\begin{equation}
\label{hneq}
H^A |A; nl; \mu\rangle = E^A_{nl}(\mu) |A; nl; \mu\rangle,
\end{equation}
where $l$ is the orbital angular momentum and $n$ (0, 1, \mbox{\ldots)}
the vibrational quantum number.
The physical solutions are obtained by choosing for each eigenstate the
value of $\mu$ which minimizes its energy.
We will note $\mu_{nl}$ the value of $\mu$ minimizing the
eigenvalue $E^A_{nl}(\mu)$. The following notations will be used
\begin{equation}
\label{ennl}
E^A_{nl} = E^A_{nl}(\mu_{nl}) \leq E^A_{nl}(\mu) \quad \forall
\mu \quad \text{and} \quad |A; nl \rangle = |A; nl; \mu_{nl}\rangle.
\end{equation}
Applying the Hellmann-Feynman theorem \cite{feyn39} to $H^A$, we find
\begin{equation}
\label{hftheo1}
\frac{\partial E^A_{nl}(\mu)}{\partial \mu} =
- \frac{\langle A; nl; \mu| \vec p\,^2 + m^2 |A; nl; \mu\rangle}{\mu^2}
+ 1.
\end{equation}
It follows immediately that
\begin{equation}
\label{hftheo2}
\mu_{nl}^2 = \langle A; nl| \vec p\,^2 + m^2 |A; nl \rangle,
\end{equation}
and that
\begin{equation}
\label{hftheo3}
E^A_{nl} = 2\, \mu_{nl} + \langle A; nl| V |A; nl \rangle.
\end{equation}
Applied to $H^A$, the virial theorem yields
$2\langle \vec p\,^2 \rangle / \mu = \langle \vec r \cdot \vec \nabla V
\rangle$. If the potential V is a homogeneous function of $r$ of degree
$\alpha$, the formula~(\ref{hftheo3}) can be simplified. We have then
\begin{equation}
\label{hftheo4}
E^A_{nl} = \frac{2 (1+\alpha)}{\alpha} \mu_{nl} -
\frac{2\ m^2}{\alpha \, \mu_{nl}} \quad \text{if} \quad \vec r \cdot
\vec \nabla V = \alpha\, V.
\end{equation}

The scalar product $\langle A; n'l'| A; nl \rangle$ is proportional to
$\delta_{l'l}$ because of the orthogonality of the spherical harmonics.
But, for a fixed value of $l$, two eigenstates with different values of
$n$ are not orthogonal since they are characterized by different values
of $\mu$. So, it is interesting to compute the quantity which measures
the overlap of two states
\begin{equation}
\label{pnnl}
P_{n'nl} = |\langle A; n'l| A; nl \rangle|^2.
\end{equation}
We have obviously $P_{n\,n'\,l}=P_{n'\,n\,l}$ and $P_{n\,n\,l}=1$, and
we can expect that $P_{n'nl} \sim \delta_{n'n}$.

Another way of dealing with auxiliary field is to minimize not the
eigenvalues but the hamiltonian operator itself and impose the condition
$\partial H^A/\partial \mu = 0$, $\mu$ being considered as an operator
by its own. In the case of hamiltonian~(\ref{hn}), it is easy to
determine that the value of the auxiliary field $\mu$ which realizes
this condition is $\mu = \sqrt{\vec p\,^2 + m^2}$. The minimal
hamiltonian~(\ref{hn}) then becomes
\begin{equation}
\label{hs}
H^S=2\sqrt{\vec p\,^2 + m^2} +V,
\end{equation}
which is simply a spinless Salpeter hamiltonian with the same potential
$V$. We note its eigenvalues and eigenstates by
\begin{equation}
\label{hseq}
H^S |S; nl\rangle = E^S_{nl} |S; nl\rangle.
\end{equation}
It has been shown in Ref.~\cite{luch96}, that we always have
$E^A_{nl} \geq  E^S_{nl}$.

If the two ways of treating the auxiliary field were expected to be very
similar, one sees that the first method simulates the effect of a
relativistic kinetic energy term with a nonrelativistic one; the
potential is kept unchanged. The differences $E^A-E^S$ give an
indication on the quality of this procedure.

Since the minimal energy eigenvalues of $H^A$ are upper bound of the
eigenvalues of $H^S$ and since these two hamiltonians have in some
sense the same physical content, their corresponding eigenstates must
be very similar. By computing the overlap
\begin{equation}
\label{tnl}
T_{nl} = |\langle A; nl| S; nl \rangle|^2,
\end{equation}
we can expect that $T_{nl} \sim 1$.

If we define
\begin{equation}
\label{Mnl}
M_{nl} = \langle S; nl | \sqrt{\vec p\,^2 + m^2} |S; nl\rangle,
\end{equation}
we can also expect that $M_{nl} \approx \mu_{nl}$, because of the
relation~(\ref{hftheo2}). In both models, $m$ is sometimes considered as
the current mass, and $M_{nl}$ or $\mu_{nl}$ are then interpreted as the
constituent mass, which, in this formalism, is state dependent.

For two massless particles, the relativistic virial theorem
\cite{luch90} yields
$2\langle \sqrt{\vec p\,^2} \rangle = \langle \vec r \cdot \vec \nabla V
\rangle$. If the potential V is a homogeneous function of $r$ of degree
$\alpha$, new relations can be written. We have then
\begin{equation}
\label{virial}
E^S_{nl} = (1+\alpha) \,\langle S; nl | V |S; nl\rangle =
\frac{2 (1+\alpha)}{\alpha} M_{nl} \quad \text{if} \quad \vec
r \cdot \vec \nabla V = \alpha\, V \quad \text{and} \quad m=0.
\end{equation}

When the mass $m$
increases, the differences existing between the two
approaches $H^A$ and $H^S$ tend to vanish and $\mu_{nl} \approx m$ for
all values of quantum numbers. In order to quantify these differences,
we will study two models containing a confining potential, which is the
relevant kind of interaction for the hadronic physics. In particular, we
will focus our attention on the case $m=0$, for which we expect the
largest discrepancies between the two approaches.

\subsection{Toy model}
\label{ssec:toy}

We first consider a ``toy model" for which most of the calculations are
analytical. Let us choose $V=k\,r^2$ and $m=0$. In this case, $H^A$
is a harmonic oscillator hamiltonian. Its solutions are well-known and
we have
\begin{equation}
\label{tenlmu}
E^A_{nl}(\mu) = 2\sqrt{\frac{k}{\mu}} (2 n + l + 3/2) + \mu.
\end{equation}
The regularized radial part of the wave function is noted
\begin{equation}
\label{tun}
u^A_{nl}(r) = (\mu k)^{3/8} \, r \, O_{nl}
\left( (\mu k)^{1/4} r \right),
\end{equation}
where $O_{nl}(x)$ is the usual radial harmonic oscillator wave function
and where dimensioned factors ensure that the eigenstates
$|A; nl;\mu\rangle$ are normalized.

The values of the parameter $\mu$ minimizing the energies are given by
\begin{equation}
\label{tmunl}
\mu_{nl}=k^{1/3} (2 n + l + 3/2)^{2/3},
\end{equation}
and the calculation of the eigenenergies yields
\begin{equation}
\label{tennl}
E^A_{nl} = 3\, k^{1/3} (2 n + l + 3/2)^{2/3} = 3\, \mu_{nl}.
\end{equation}
The relation $E^A_{nl} = 3\, \mu_{nl}$ can also be obtained directly
from formula~(\ref{hftheo4}).

The overlap of two eigenstates can be computed easily. Knowing the
values of $\mu_{nl}$, we have
\begin{equation}
\label{tpnnl}
P_{n'nl}=F_{n'nl}^2\left( \left(\frac{2n+l+3/2}{2n'+l+3/2}\right)^{1/6}
\right),
\end{equation}
where the function $F_{n'nl}(x)$ is given, with its properties, in
Ref.~\cite{sema95a}. The quantity $P_{n'nl}$ is independent of $k$ and
decreases with increasing values of $l$ and $|n-n'|$. For instance, we
have $P_{1\,0\,0} \approx 0.0287$, which is the worse case. Considering
the states as orthogonal is thus always a good approximation.

In order to solve the corresponding hamiltonian $H^S$ for our toy model,
let us consider the following nonrelativistic hamiltonian
\begin{equation}
\label{thl}
H^N=\frac{\vec p\,^2}{2\, \nu} + a\, r.
\end{equation}
Solutions for $l=0$ are known \cite{luch91}. The eigenenergies are given
by
\begin{equation}
\label{tel}
E^N_{n0} = -\left( \frac{a^2}{2\, \nu} \right)^{1/3} x_n,
\end{equation}
where $x_n$ is the $(n+1)$th zero of the Airy function $\text{Ai}(x)$
($x_0\approx -2.338$, $x_1\approx -4.088$, \mbox{\ldots)}. The
regularized and normalized radial part of the eigenvectors is written
as \cite{brau98}
\begin{equation}
\label{tul}
u^N_{n0}(r) = (2\, \nu\, a)^{1/6} A_n \left((2\, \nu\,a)^{1/3}\,
r \right) \quad \text{where} \quad A_n(x) =
\frac{\text{Ai}(x+x_n)}{\sqrt{\int_{x_n}^\infty \text{Ai}^2(q)\,dq}}.
\end{equation}
Let us note that we have
$\int_{x_n}^\infty \text{Ai}^2(q)\,dq={\text{Ai}'}^2(x_n)$.

The hamiltonian $H^N$ is converted into the hamiltonian $H^S$ with
$V=k\,r^2$ and $m=0$ by the duality transformation
$|\vec p\,| \longleftrightarrow \displaystyle\frac{1}{2} a\, r$,
provided the parameters $k$, $a$, and $\nu$ are related by
$k=\displaystyle\frac{a^2}{8\, \nu}$ \cite{luch91}. Due to this
relation, we have immediately the property
\begin{equation}
\label{tes}
E^S_{n0}=-(4\, k)^{1/3}\, x_n.
\end{equation}
In order to compare energies $E^A_{n0}$ and $E^S_{n0}$ given
respectively by relations~(\ref{tennl}) and (\ref{tes}), we can use the
following approximation for the numbers $x_n$ \cite{abra70}
\begin{equation}
\label{tzeroai}
x_n \approx - \left[ \frac{3\pi}{2} ( n + 3/4 )
\right]^{2/3}.
\end{equation}
With this formula, whose error decreases when $n$ increases,
relation~(\ref{tes}) can be written
\begin{equation}
\label{tesapp}
E^S_{n0} \approx \left( \frac{3\pi}{2} \right)^{2/3} k^{1/3}
(2 n + 3/2)^{2/3}.
\end{equation}
So we have immediately the ratio (independent of $k$),
\begin{equation}
\label{tesrat}
\frac{E^S_{n0}}{E^A_{n0}} \approx \frac{1}{3} \left( \frac{3\pi}{2}
\right)^{2/3} \approx 0.937.
\end{equation}
Numerically, we found that the ratio $E^S_{n0}/E^A_{n0}$ is around
0.944 for $n=0$ and tends rapidly towards the asymptotical
value~(\ref{tesrat}) as $n$ increases.

From relation~(\ref{tul}) and the duality transformation, the
regularized function of S-wave eigenvectors of $H^S$ can be obtained in
the momentum space
\begin{equation}
\label{tus}
u^S_{n0}(p) = \left(\frac{2}{k}\right)^{1/6} A_n
\left( \left( \frac{2}{k} \right)^{1/3} p \right).
\end{equation}
Consequently, the quantity
$M_{n0}=\langle S; nl | \sqrt{\vec p\,^2} |S; nl\rangle$ can be
calculated by the following
integral ($\sqrt{\vec p\,^2} = |\vec p\,|$)
\begin{equation}
\label{tmnl}
M_{n0} = k^{1/3} \frac{1}{2^{1/3}} \int_0^\infty dq\, q\,
A_n^2(q) = - (4\, k)^{1/3} \frac{x_n}{3} = \frac{E^S_{n0}}{3},
\end{equation}
and we find that $M_{n0}/\mu_{n0} = E^S_{n0}/E^A_{n0} \approx 0.937$.
The relation $E^S_{n0} = 3\, M_{n0}$ can also be obtained directly from
formula~(\ref{virial}).

Since the Fourier transform of a harmonic oscillator is also a harmonic
oscillator (with a phase factor), the overlap $T_{n0}$ can be computed
in the momentum space. We obtain
\begin{equation}
\label{ttn}
T_{n0} = \frac{2^{1/3}}{\sqrt{2 n +3/2}} \left[
\int_0^\infty dq\, q\,
O_{n0} \left( \frac{q}{(2n+3/2)^{1/6}}  \right)
A_n \left( 2^{1/3} q \right)
\right]^2,
\end{equation}
which is independent of $k$.
We have checked numerically that $T_{0\,0}\approx 0.997$ and that
$T_{n0}$ decreases monotonically when $n$ increases. For instance,
$T_{9\,0}\approx 0.523$. One sees that, when $n$ increases, the
eigenvalues for the two formalisms are almost the same (to within 6\%),
whereas the corresponding wave functions can differ appreciably.

\subsection{Realistic model}
\label{ssec:real}

The previous toy model was interesting because most of the quantities
can be calculated analytically. However, it lacks some physics. First
the real confining potential is more alike a linear one, and second
there exist one gluon exchange contributions which add a coulombic term,
important at short distance. In order to stick more to the true physical
situation, we switch, in this section, to a more realistic model. Of
course, this study needs a complete numerical treatment.

In the framework of a semirelativistic potential model, a more realistic
interaction to simulate the dynamics between a quark and an
antiquark inside a meson is certainly what is called the funnel
potential \cite{fulc94}
\begin{equation}
\label{vreal}
V(r) = -\frac{\kappa}{r} + a r \quad \text{with} \quad \kappa=0.5 \quad
\text{and} \quad a=0.2\ \text{GeV}^2.
\end{equation}
The values chosen here for the parameters $\kappa$ and $a$ can be
considered as typical.
The eigenvalue equations for hamiltonians $H^A$ and $H^S$ with this
potential have been numerically solved by the Lagrange-mesh method
\cite{sema01}. This technique is very accurate and can be implemented
easily for both nonrelativistic or semirelativistic kinematics.

\begin{center}
\begin{figure}
\includegraphics*[height=8cm]{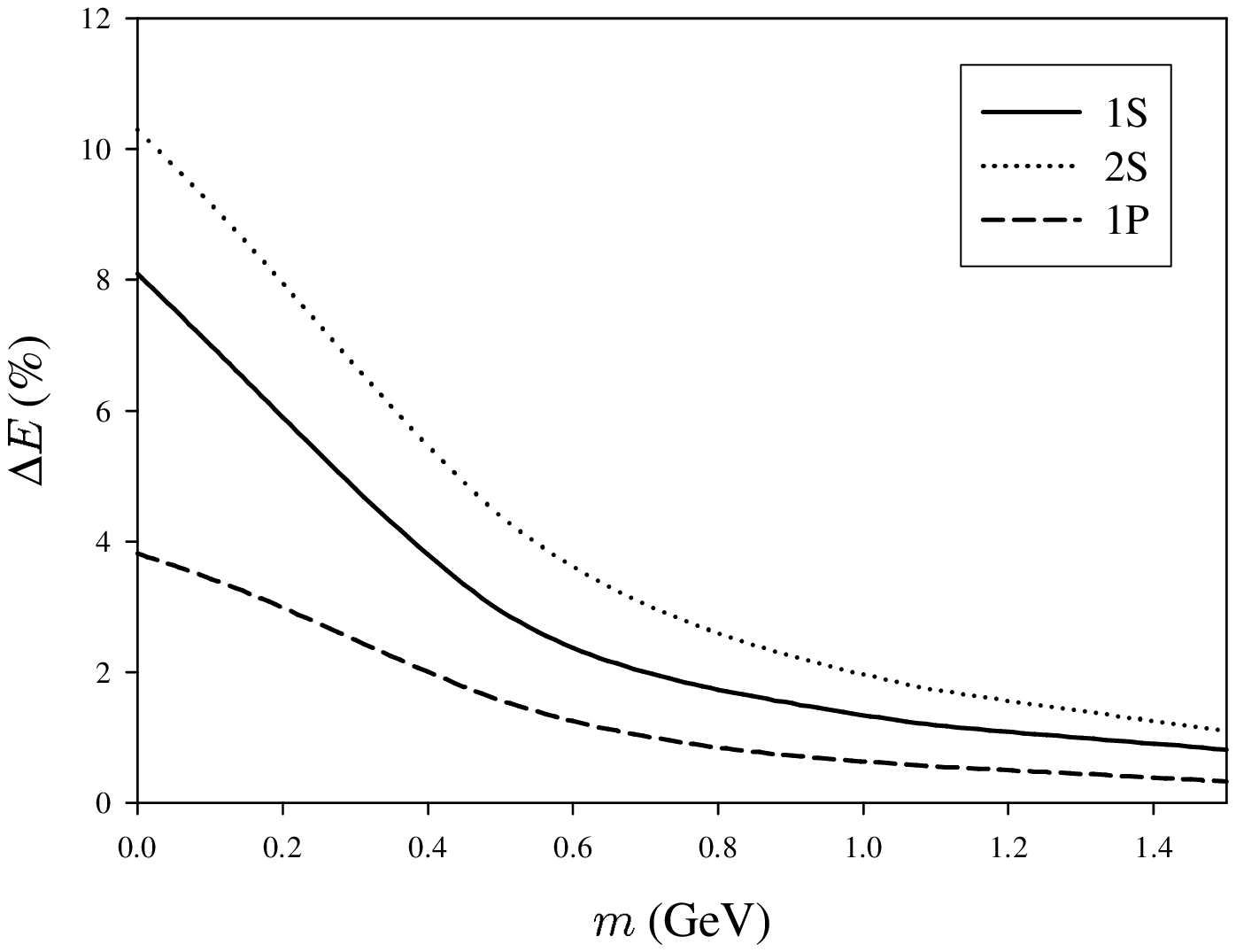}
% 886 x 716
%\centerbmp{9.90cm}{8cm}{fig1.bmp}
\caption{$\Delta E = (E^A - E^S)/E^S$ (see Sec.~\ref{sec:sse}) as a
function of the mass $m$, for 1S, 2S, and 1P states with
potential~(\ref{vreal}).}
\label{fig1}
\end{figure}
\end{center}

In Fig.~\ref{fig1}, the quantities $\Delta E = (E^A - E^S)/E^S$ are
presented, as a function of the
mass $m$, for the three lowest states (1S, 2S, 1P) obtained with the
potential~(\ref{vreal}). As expected, $\Delta E$ decreases when the
quark mass increases. The corresponding curves for the quantities
$E^A - E^S$ present the same profiles. Thus we only present more results
for the case of a vanishing quark mass which maximizes the error.

In Table~\ref{tab1}, some energies $E^S_{nl}$ and $E^A_{nl}$ are given
for $n \leq 3$ and $l \leq 2$, for the
potential~(\ref{vreal}) and $m=0$. The difference $E^A_{nl} - E^S_{nl}$
can be quite large, around 100~MeV for the ground state. It increases
for increasing values of $n$ and decreases for increasing values of $l$.
The quantity $M_{nl}$ is also given in this Table; it increases with $n$
and $l$. The parameter $\mu_{nl}$ can differ from $M_{nl}$ by several
tens of MeV, but the curves $E^A_{nl}(\mu)$ always present a flat
minimum around $\mu_{nl}$. So it is not necessary to obtain a precise
value of $\mu_{nl}$ to obtain a good value of $E^A_{nl}$.

The overlap $P_{n\,n'\,l}$ is given in Table~\ref{tab2} for
$n, n' \leq 3$ and for $l \leq 1$, for the potential~(\ref{vreal}) and
$m=0$. Except in the case $|n-n'|=1$, the overlap is very small. So the
eigenstates $|A; nl \rangle$ can be considered as quasi-orthogonal. This
conclusion is the same than in the toy model, and we can hope that it is
a universal conclusion, valid whatever the potential $V$.

In table~\ref{tab3}, the overlap $T_{n\,l}$ is presented for $n \leq 3$
and $l \leq 2$, for the potential~(\ref{vreal}) and $m=0$. It increases
for increasing values of $l$ and decreases for increasing values of $n$,
and present similar features to the toy model.
For small values of the vibrational quantum number, the eigenvectors
$|A; nl \rangle$ and $|S; nl \rangle$ are rather similar.

\section{Rotating QCD string}
\label{sec:string}

Up to now, the confining term was put by hand. There exist more serious
explanations of this term, based on QCD arguments. In particular, one
can consider a string with an energy density between the quark and the
antiquark. This string is itself a dynamical object. Here we examine two
such approaches.

We first quickly present the relativistic flux tube model, which is
essentially phenomenological, and then the rotating string model within
the formalism of auxiliary fields, which has a more firm foundation on
QCD theory. Lastly, we show that these two models are completely
equivalent.

\subsection{Phenomenological model}
\label{ssec:flux}

In the simplest version of the relativistic flux tube model
\cite{laco89}, a quark and an antiquark, with the same mass $m$, move
being attached with a rigid flux tube, assumed to be linear with a
uniform constant energy density $a$. The system rotates in a plane with
a constant angular velocity around the center of mass, which is assumed
to be stationary. Denoting $r$ the distance between the two particles,
$p_r$ the radial momentum
$\left( p_r^2 = -\displaystyle\frac{1}{r}
\displaystyle\frac{\partial^2}{\partial r ^2} r \right)$, $L$ the total
angular momentum, $v_\perp$ the transverse velocity of the quark or the
antiquark (relative to the string direction),
$\gamma_\perp=(1-v_\perp^2)^{-1/2}$, and $U(r)$ a potential used to
describe dynamical effects coming from mechanisms other than the flux
tube, the classical equations of motion of the system are given by \cite
{laco89}
\begin{eqnarray}
\label{hft1}
\frac{L}{r} &=& v_\perp \gamma_\perp \sqrt{p_r^2 + m^2} +
a\, r\,f(v_\perp), \\
\label{hft2}
H &=& 2 \gamma_\perp \sqrt{p_r^2 + m^2} + a\, r
\frac{\arcsin v_\perp}{v_\perp} + U(r),
\end{eqnarray}
where
\begin{equation}
\label{ffx}
f(x)=\frac{1}{4\,x^2} \left(\arcsin x -x\sqrt{1-x^2}\right),
\end{equation}
is a very important function in the formalism.
These equations have been generalized in the case of asymmetrical
systems \cite{olss95}. Numerical solutions of the quantized versions of
these equations have been obtained \cite{laco89,sema94,olss95,sema95b}
(for a practical calculation $L$ must be replaced by $\sqrt{l(l+1)}$).
Equations~(\ref{hft1})-(\ref{hft2}) are coupled nonlinear equations.
In practice, the value $v_\perp(m,a,L;p_r,r)$ is extracted from
Eq.~(\ref{hft1}) and injected into Eq.~(\ref{hft2}), giving a $L$
dependent hamiltonian $H(m,a,L;p_r,r)$, which is diagonalized
afterwards.

When the relativistic flux tube hamiltonian is supplemented by
appropriate potentials (Coulomb-like, instanton induced effects) and
when it is assumed that each extremity of the flux tube can give a
constant energy contribution, rather good meson spectra
can be obtained \cite{sema94,sema95b}.

\subsection{Auxiliary fields for a rotating string}
\label{ssec:string}

Starting from the QCD lagrangian, a lagrangian for a meson can be
derived taking into account the dynamical degrees of freedom of the
string \cite{morg99}. For a system containing a quark and an antiquark
with the same mass, with the hypothesis of a straight line configuration
for the minimal string, and introducing an auxiliary field $\mu$ for the
energy density of the quarks and another auxiliary field $\nu$ for the
energy density of the string, the following classical hamiltonian can be
obtained \cite{morg99}
\begin{eqnarray}
\label{roth}
H=\frac{p_r^2 + m^2}{\mu(\tau)} + \mu(\tau) &+&
\frac{L^2/r^2}{\mu(\tau)+2\displaystyle\int_0^1 \left( \beta-\frac{1}{2}
\right)^2
\nu(\beta,\tau)\, d\beta} \nonumber \\
&+& \frac{a^2 r^2}{2} \int_0^1
\frac{d\beta}{\nu(\beta,\tau)} + \int_0^1 \frac{\nu(\beta,\tau)}{2}
d\beta + U(r),
\end{eqnarray}
where $U(r)$ is a potential describing dynamical effects
coming from mechanisms other than the string. In this formula,
$\tau$ is the common proper time of the two particles and
$\beta$ is a coordinate along the string. Within this formalism, the two
auxiliary functions $\mu$ and $\nu$ are to be varied and to be found
from the minimum of $H$.

Let us first remark that, if we neglect the integral in the denominator
of the term depending on $L^2$ or if we put $L=0$, then $H$ is minimum
for $\nu = a\,r$, independent of $\beta$. In this case, because of the
relation $p_r^2+\displaystyle\frac{L^2}{r^2}=\vec p\,^2$, $H$ reduces
simply to $H^A$ with $V(r)=a\,r+U(r)$. The linear confining term appears
naturally in the formalism. A further minimization on $\mu$, as
demonstrated in second section, gives rise to $H^S$ with the same
potential $V(r)$.

This result is interesting and simple, but indeed one can make an exact
minimization on the field $\nu$. A straightforwards calculation
shows that the minimization condition $\partial H/\partial \nu = 0$ is
fulfilled if the field $\nu$ is set to $\nu_0$ with
\begin{equation}
\label{nub0}
\nu_0(\beta) = \frac{a\, r}{\sqrt{1-4\, y^2 \left( \beta-\frac{1}{2}
\right)^2}},
\end{equation}
where $y$ is to be found from the transcendental equation
\begin{equation}
\label{L}
\frac{L}{a\, r^2} = f(y) + \frac{\mu\, y}{a\, r}.
\end{equation}
The function $f$ defined previously by relation~(\ref{ffx}) appears
again curiously (for a practical calculation $L$ must also be replaced
by $\sqrt{l(l+1)}$). Using expression~(\ref{nub0}), one obtains from
relation~(\ref{roth})
\begin{equation}
\label{hmin}
H=\frac{p_r^2 + m^2}{\mu} + \mu (1 + y^2) +
a\, r\frac{\arcsin y}{y} + U(r).
\end{equation}

WKB solutions for the system~(\ref{L}) and (\ref{hmin}) are obtained in
Ref.~\cite{morg99} for $m=0$ and $U(r)=0$. Regge trajectories are
computed in agreement with experimental data.

\subsection{Equivalence between the two string models}
\label{ssec:equiv}

In principle the way of solving the system~(\ref{L})-(\ref{hmin}) is the
following. One extracts $y(a,L;r;\mu)$ from Eq.~(\ref{L}) and inject it
into Eq.~(\ref{hmin}) leading to $H(m,a,L;p_r,r;\mu)$. This hamiltonian
is then diagonalized and, for each state, the optimal value of $\mu$
must be determined to get the physical eigenenergies.
In view of the form for various expressions, this
procedure seems hopeless. Fortunately this is not so and one can go one
step further and perform the minimization on $\mu$.

The partial derivative of Eq.~(\ref{L}) with respect to $\mu$ gives the
following relation
\begin{equation}
\label{der1}
\frac{y^2}{a\, r} + \frac{\partial y}{\partial \mu} \left[
\frac{\mu\, y}{a\, r} + \frac{1}{2\, y \sqrt{1-y^2}} -
\frac{\arcsin y}{2\, y^2} \right] = 0,
\end{equation}
in which the expression $\partial y/\partial \mu$ appears.
Now imposing $\partial H/\partial \mu = 0$, we obtain from
Eq.~(\ref{hmin})
\begin{equation}
\label{der2}
-\frac{p_r^2 + m^2}{\mu^2} + 1 + y^2 + \frac{\partial y}{\partial \mu}
\left[ 2\, \mu\, y + \frac{a\, r}{y \sqrt{1-y^2}} - a\, r
\frac{\arcsin y}{y^2} \right] = 0.
\end{equation}
Combining the two last formulas, we obtain
\begin{equation}
\label{prmu}
\mu = \sqrt{\frac{p_r^2 + m^2}{1-y^2}}.
\end{equation}
Using this relation, it can be verified that Eqs.~(\ref{der1}) and
(\ref{der2}) are well defined for $y \in [0,1]$.

%Now the problem is simplified. Equation~(\ref{prmu}) gives $\mu(y)$
%which is used in Eq.~(\ref{L}) to get directly $y(m,a,L;p_r,r)$ and
%then $\mu(m,a,L;p_r,r)$. Inserting those functions into
%Eq.~(\ref{hmin}) provides directly $H(m,a,L;p_r,r)$, which must be
%diagonalized. One sees that this procedure is very similar to the one
%needed in the relativistic flux tube model.

Equation~(\ref{prmu}) giving $\mu(y)$ and the presence of the function
$f$~(\ref{ffx}) in Eq.~(\ref{L}) are the clues for the identification of
both methods. Replacing $y$ by $v_\perp$ and $\mu$ by
$\gamma_\perp \sqrt{p_r^2+m^2}$ transforms Eq.~(\ref{L}) into
Eq.~(\ref{hft1}) and Eq.~(\ref{hmin}) into Eq.~(\ref{hft2}), proving
thus the complete equivalence between both approaches. In passing we
have now a clear idea of the physical content for the auxiliary field
$\mu$ and of the mysterious variable $y$ appearing in the theory.

\section{Conclusion}
\label{sec:conc}

Effective hamiltonians for hadron dynamics can be derived from the
original QCD lagrangian. These hamiltonians can depend on auxiliary
fields, representing for instance the quark energy density
\cite{kala97,naro02}. When the interaction does not depend on these
fields (exactly or by approximation), it is possible to reduce the
effective hamiltonian to a semirelativistic one (spinless Salpeter).

In the first part of this paper, we have compared the eigenvalues and
eigenvectors for hamiltonians written in the auxiliary field formalism
$H^A$ and for the corresponding spinless Salpeter hamiltonians $H^S$. We
have only considered the case of two particles with the same mass and
two different confining interaction: A ``toy" quadratic potential and a
more realistic funnel potential. We have shown that the eigenvalues of
$H^A$ are close to the eigenvalues of $H^S$, but the relative
differences can exceed 10\%. The overlap of corresponding eigenvectors
for the two approaches is generally close to unity for ground states,
but it deteriorates when the vibrational quantum number increases. It
appears that, if precise calculations are desired, it is better to solve
directly a spinless Salpeter equations: Accurate numerical techniques
exist for two-body \cite{sema01}, and many-body \cite{suzu98,silv01}
problems. Hamiltonians with auxiliary fields can be useful if one
searches to obtain analytical results, as it is the case in
Ref.~\cite{naro02}. The non-orthogonality resulting in this formalism is
not a serious problem, the radial excited states being quasi-orthogonal.

The phenomenological semirelativistic flux tube hamiltonian has been
developed in order to take into account the string dynamics in a meson
\cite{laco89,olss95}. Another model derived from the QCD lagrangian, the
rotating string, has been more recently developed \cite{morg99}. But the
rotating string hamiltonian depends on auxiliary fields.

In the second part of this paper, we have shown that the classical
equations of motion of the rotating string reduce exactly to the
classical equations of motion of the semirelativistic flux tube,
provided all auxiliary fields are correctly eliminated. The complete
equivalence of both models has two virtues: to provide a natural
interpretation of the auxiliary field in term of the transverse velocity
and to reinforce the relevance of the relativistic flux tube model
(which was considered up to now as a phenomenological one), since it is
equivalent to a model based on firm QCD grounds.

The set of coupled equations for these two models are very complicated
to solve numerically \cite{laco89,sema94,olss95,sema95b}. The
difficulties are essentially the same for both approaches, with the
supplementary necessity to perform a minimization within the rotating
string formalism in order to determine the value of the auxiliary field
for the quark energy density. It is then more interesting to work
directly with the semirelativistic flux tube model.

\section{Acknowledgments}

The authors are grateful to NATO services which give us the possibility
to visit ourselves and greatly facilitates this work, through the NATO
grant PST.CLG.978710. B. Silvestre-Brac and C. Semay (FNRS Research
Associate position) would like to thank the agreement CNRS/CGRI-FNRS for
financial support. They also thank A. M. Badalian, F. Brau, Yu. S.
Kalashnikova, and Yu. A. Simonov for useful discussions.

%\appendix

\clearpage

\begin{table}
\protect\caption{Values in GeV for the quantities $E^S_{nl}$, $E^A_{nl}$
and $M_{nl}$
(see Sec.~\ref{sec:sse}), as a function of the quantum numbers $n$ and
$l$, for the potential~(\ref{vreal}) and $m=0$.}
\label{tab1}
\begin{ruledtabular}
\begin{tabular}{cccccc}
 & $n$ & 0 & 1 & 2 & 3 \\
\hline
$l=0$ & $E^S_{nl}$ & 1.197 & 1.912 & 2.461 & 2.914 \\
      & $E^A_{nl}$ & 1.294 & 2.109 & 2.700 & 3.185 \\
      & $M_{nl}$   & 0.422 & 0.588 & 0.712 & 0.817 \\
\hline
$l=1$ & $E^S_{nl}$ & 1.759 & 2.316 & 2.782 & 3.187 \\
      & $E^A_{nl}$ & 1.826 & 2.472 & 2.990 & 3.434 \\
      & $M_{nl}$   & 0.508 & 0.644 & 0.757 & 0.855 \\
\hline
$l=2$ & $E^S_{nl}$ & 2.170 & 2.643 & 3.056 & 3.426 \\
      & $E^A_{nl}$ & 2.225 & 2.780 & 3.248 & 3.659 \\
      & $M_{nl}$   & 0.594 & 0.711 & 0.813 & 0.904 \\
\end{tabular}
\end{ruledtabular}
\end{table}

\begin{table}
\protect\caption{Values of $P_{n\,n'\,l}$ (see Sec.~\ref{sec:sse}), as a
function of the quantum numbers $n$, $n'$, and $l$, for the
potential~(\ref{vreal}) and $m=0$. Values for $l=0$ ($l=1$)
are given in the upper-right (lower-left) triangle of the Table
($P_{n\,n'\,l}=P_{n'\,n\,l}$ and $P_{n\,n\,l}=1$).}
\label{tab2}
%\begin{ruledtabular}
\begin{tabular}{r|cccc}
$P_{n\,n'\,l}$ & $n=0$ & 1 & 2 & 3 \\
\hline
$n'=0$ & 1 & 0.022 & $1.6\ 10^{-4}$ & $2.4\ 10^{-4}$ \\
1 &  0.017 & 1 & 0.022 & $5.7\ 10^{-5}$ \\
2 & $2.0\ 10^{-6}$ & 0.023 & 1 & 0.028 \\
3 & $5.9\ 10^{-5}$ & $2.5\ 10^{-7}$ & 0.019 & 1
\end{tabular}
%\end{ruledtabular}
\end{table}

\begin{table}
\protect\caption{Values of $T_{n\,l}$ (see Sec.~\ref{sec:sse}), as a
function of the quantum numbers $n$ and $l$, for the
potential~(\ref{vreal}) and $m=0$.}
\label{tab3}
%\begin{ruledtabular}
\begin{tabular}{r|cccc}
$T_{n\,l}$ & $n=0$ & 1 & 2 & 3 \\
\hline
$l=0$ & 0.9837 & 0.9639 & 0.9389 & 0.8982 \\
1     & 0.9908 & 0.9722 & 0.9406 & 0.9005 \\
2     & 0.9925 & 0.9746 & 0.9443 & 0.9008 \\
\end{tabular}
%\end{ruledtabular}
\end{table}

\clearpage

\end{document}